\documentclass{ws-ijmpcs}

\newcommand{\Tensor}[1] {\mathbb{#1}}
\newcommand{\TensorT} {\mathcal{T\!\!\!\!T}}
\newcommand{\beq}{\begin{equation}}
\newcommand{\eeq}{\end{equation}}
 \newcommand{\Label}[1]{\label{#1}}
\newcommand{\mean}[1]{\left\langle #1\right\rangle}

\begin{document}

\markboth{P.~Rodriguez-Lopez, R.~Brito and R.~Soto}
{Stochastic Quantization and Casimir Forces: Pistons of Arbitrary Cross Section}

%
\catchline{}{}{}{}{}
%

\title{Stochastic Quantization and Casimir Forces: Pistons of Arbitrary Cross Section}

\author{Pablo Rodriguez-Lopez \and  Ricardo Brito}

\address{Departamento de F\'\i sica Aplicada I and GISC, 
Universidad Complutense, 28040 Madrid, Spain\\
brito@fis.ucm.es}

\author{Rodrigo Soto}

\address{Departamento de F\'\i sica, FCFM, Universidad de Chile, Casilla 487-3, Santiago, Chile\\
rsoto@dfi.uchile.cl}
\maketitle

\begin{history}
\received{Day Month Year}
\revised{Day Month Year}
\end{history}

\begin{abstract}
Recently, a method based on stochastic quantization has been proposed to compute the Casimir force and its fluctuations in arbitrary geometries. It relies on the spectral decomposition of the Laplacian operator in the given geometry. Both quantum and thermal fluctuations are considered. Here we use such method to compute the Casimir force on the plates of a
finite piston of arbitrary cross section. Asymptotic expressions  valid at low and high temperatures and short and long distances are obtained. The case of a piston with triangular cross section is analyzed in detail. The regularization of the divergent stress tensor is described.
\end{abstract}

\ccode{PACS numbers: 11.25.Hf, 123.1K}

\section{Introduction}
H.G. Casimir~\cite{Casimir_Placas_Paralelas} predicted in 1948 the existence of an attactive force 
between two perfect parallel conducting plates due to  quantum fluctuations of the vacuum~\cite{Review_Casimir}.
Advanced  experimental setups allow to measure such 
forces  for geometries other than the two parallel plates~\cite{review2}. 
However, techniques to calculate forces for complicated geometries,
beyond the usual ones with high degrees of symmetry, were scarce until 
Ref.~\refcite{Kardar} introduced  a method  to calculate electromagnetic (EM)
Casimir forces. It is based
on a multiscattering calculation  and has been successfully applied to
many configurations, such as plates, cylinders, spheres, wedges, etc.~\cite{Casimir_Wedges}

Casimir forces have their origin in fluctuations of EM fields.
Already Lifshitz in 1956 \cite{Lifshitz} use the concept of
fluctuations (in this case of the electric currents inside a metal) to 
compute such forces. In Ref. \refcite{ParisiWuCasimir} a new method to compute Casimir forces has been derived by us. It is based on the stochastic quantization approach 
developed by  Parisi and Wu~\cite{Parisi-Wu_original}. A Langevin equation
for a given field subjected to thermal fluctuations, is constructed in a non-physical, pseudo-time, such that it reproduces the probability distribution for the field in a thermal bath. Using the formalism developed in Ref.~\refcite{We-PRE} to compute fluctuation-induced forces starting from a stochastic equation, the Casimir force of the electromagnetic field can be  obtained. The resulting method has the advantage of computing the Casimir force
directly and can be  applied to compute torques and stresses
over extended bodies as well. Also, the force fluctuations are obtained properly regularized and it was shown that the force variance on a piston geometry is finite, and twice the square of the force~\cite{ParisiWuCasimir}. 

In this manuscript we apply this method to the computation of the force on pistons of arbitrary cross section. The text is organized as follows.   
Section \ref{sec.formalism} presents a summary of the procedure to obtain the Casimir force using the stochastic quantization method. An analysis of the divergence of the stress tensor is performed and a force regularization method is proposed. In Section \ref{sec.em} the method is applied to the case of the electromagnetic Casimir force on pistons. The decomposition in transverse electric and transverse magnetic modes is described resulting in an expression for the force on the plates in terms of the eigenvalues of the 2D Laplacian on the piston cross section. The force regularization is analyzed numerically and treated analytically resulting in absolutely convergent expressions. Asymptotic formulas are provided. The case of pistons with polygonal cross section is analyzed in section \ref{sec.polygons} and the use of the method is exemplified by computing the force on pistons with a triangular cross section. Finally, conclusions are in Section \ref{sec.conclusions}.

\section{General formalism} \label{sec.formalism}
 Here we summarize the formalism developed in Ref. \refcite{ParisiWuCasimir}. The Casimir force can be obtained averaging the stress tensor over the probability distribution of the quantum field. The stochastic quantization method of Parisi and Wu~\cite{Parisi-Wu_original,Masujima} provides a stochastic equation in a pseudo time, $s$, that reproduces the full probability distribution in space, ${\bf r}$, and Wick-rotated time, $\tau= i t$, of a quantum field at temperature $T$.  
In the case of a scalar bosonic field of zero mass $\phi$, such stochastic equation reads 
\beq\label{Langevin}
\partial_s\phi (\tau ,{\bf r};s)  
 =  \left(\frac{1}{c^2}\frac{\partial^2}{\partial \tau^2}+\nabla^2\right) \phi +\eta(\tau ,{\bf r};s),
\eeq
where $\eta(\tau ,{\bf r};s)$ is the noise term, mimicking the source of 
fluctuations for the field $\phi$. It is a white noise, $\delta$-correlated in $\tau$, ${\bf r}$, and $s$, with intensity 
$2k_BT$. Equation (\ref{Langevin}) can be solved  by diagonalization.
Its  temporal part involves the Matsubara frequencies $\omega_m=2\pi m/\beta\hbar c$, $m\in\mathbb{Z}$,
and the spatial part requires the diagonalization  of the Laplacian in the 
geometry with the appropriate boundary conditions (BC),
\beq
\nabla^2 f_n({\bf r})=-\lambda_n^2 f_n({\bf r}).
\label{eq.eigenfunctions}
\eeq

If the stress tensor is a bilinear function of the field (as is the case
in electromagnetism), $\Tensor{T}=\TensorT[\phi,\phi,{\bf r}]$, which defines the stress tensor operator $\TensorT$, 
then the average over the fluctuations can be computed by decomposing the field $\phi$ into its eigenfunctions,
to obtain (see details in Ref. \refcite{ParisiWuCasimir}):
\begin{equation}
\langle\Tensor{T}({\bf r})\rangle =
\frac{1}{\beta}\sum_{nm}\frac{\TensorT_{nn}({\bf r})}
{\lambda_n^2+\omega_m^2}, \label{eq.T}
\end{equation}
where $\TensorT_{nn}({\bf r})=\TensorT[f_n,f_n^*,{\bf r}]$. 
The sum over the Matsubara frequencies, $\omega_m$, gives,
\beq 
\langle \Tensor{T}({\bf
r})\rangle= \frac{\hbar c}{2}\sum_{n} \frac{\TensorT_{nn}({\bf r})}
{\lambda_n} \left[1+\frac{2}{e^{\beta \hbar c \lambda_n}-1}\right].
\Label{eq.T2}
\eeq 
Finally, to obtain the Casimir force over a~certain body, the stress
tensor must be integrated over the surface~$\Omega$ that defines the
object
\begin{equation}  \Label{eq.FCint}
{\bf F}_{C}  = \oint_\Omega \langle \Tensor{T}({\bf r})\rangle\cdot d{\bf S} . 
\end{equation}

This expression allows to compute the quantum Casimir force including the
effects of a finite, nonvanishing temperature, in terms of the eigenvalues and eigenfunctions of the
Laplacian operator.

\subsection{Regularization} \label{subsec.regularization}
Let us note that, as it occurs in most of the calculations of the Casimir Force, 
the expression for $\langle \Tensor{T}({\bf r})\rangle$ in Eq. (\ref{eq.T2}) is generally 
divergent at every point of space ${\bf r}$ if the sum over eigenvalues runs to infinity. 
Indeed, an estimation can be made considering Fourier modes with wavevector ${\bf k}$. The eigenvalues of the Laplacian grow as 
$\lambda^2\sim k^2$ and the contribution of each mode to the stress tensor in the case of Dirichlet BC (as is the case in electromagnetism) is also $\TensorT_k\sim k^2$.
Substitution into Eq.~(\ref{eq.T2})  results in a divergent expression for the stress tensor on the object boundary.

However, the expression for the force, that is obtained integrating over the surface of 
the body, Eq.(\ref{eq.FCint}), is finite. It means that the integration over the body 
regularizes the divergences of the averaged stress tensor. If we assume that 
such regularization is carried out mode by mode, we can interchange the integration over the surface 
and the summation over eigenvalues, to obtain, 
\beq 
{\bf F}_{C} = \frac{\hbar c}{2}\sum_{n} \frac{1}
{\lambda_n} \left[1+\frac{2}{e^{\beta \hbar c \lambda_n}-1}\right]
\oint_\Omega \TensorT_{nn}({\bf r}) \cdot d{\bf S},
\Label{eq.T3}
\eeq 
which is a {\em finite} result. Therefore, the interchange of the integral and summation
regularizes the Casimir force, avoiding the use of ultraviolet cutoffs. 
Other regularizations, that in some cases may lead to non-universal forces or fluctuations
are, for instance, the subtraction of the vacuum stress tensor \cite{Ford} or
by averaging the stress tensor over a finite area or a finite time \cite{Barton}.
Having regularized the divergences, Eq. (\ref{eq.T3}),
provides a new expression  to calculate Casimir forces for a~given geometry by diagonalizing the
Laplace operator. So, this approach is suitable for numerical calculations of Casimir forces
in complicated, realistic geometries. Moreover, this method leads to the force directly, not
as a difference of the free energy with respect to a reference state, which in some configurations it may be
difficult to establish. 
Other authors  have obtained formulas for the free energy in terms of the 
eigenvalues of the spatial operator, but not for the force~\cite{Marachevsky,Abalo}.

The expression for the Casimir force, Eq.~(\ref{eq.T3}), allows one
to evaluate the quantum limit (by setting the temperature
equal to zero) and the classical limit
\beq
\lim_{T\to0}{\bf F}_{C} = \frac{\hbar c}{2}\sum_{n} \frac{1}
{\lambda_n} \oint_\Omega \TensorT_{nn}({\bf r}) \cdot d{\bf S}; \quad
\lim_{\hbar \to 0}{\bf F}_{C} = k_BT\sum_{n} \frac{1}
{\lambda_n^2} \oint_\Omega \TensorT_{nn}({\bf r}) \cdot d{\bf S}.
\Label{eq.FCh0} 
 \eeq 
The convergence of these expressions depend on the sum and integral exchange. 
The two limits, quantum 
and thermal, show that the driving force of the
fluctuations has different origin. In the first case, the presence
of the factor~$\hbar$ indicates the quantum nature of the
fluctuations, whereas in the second case, the factor $k_BT$ reveals
its thermal origin.

\begin{figure}
\begin{center}\includegraphics[width=0.7\columnwidth,angle=0]{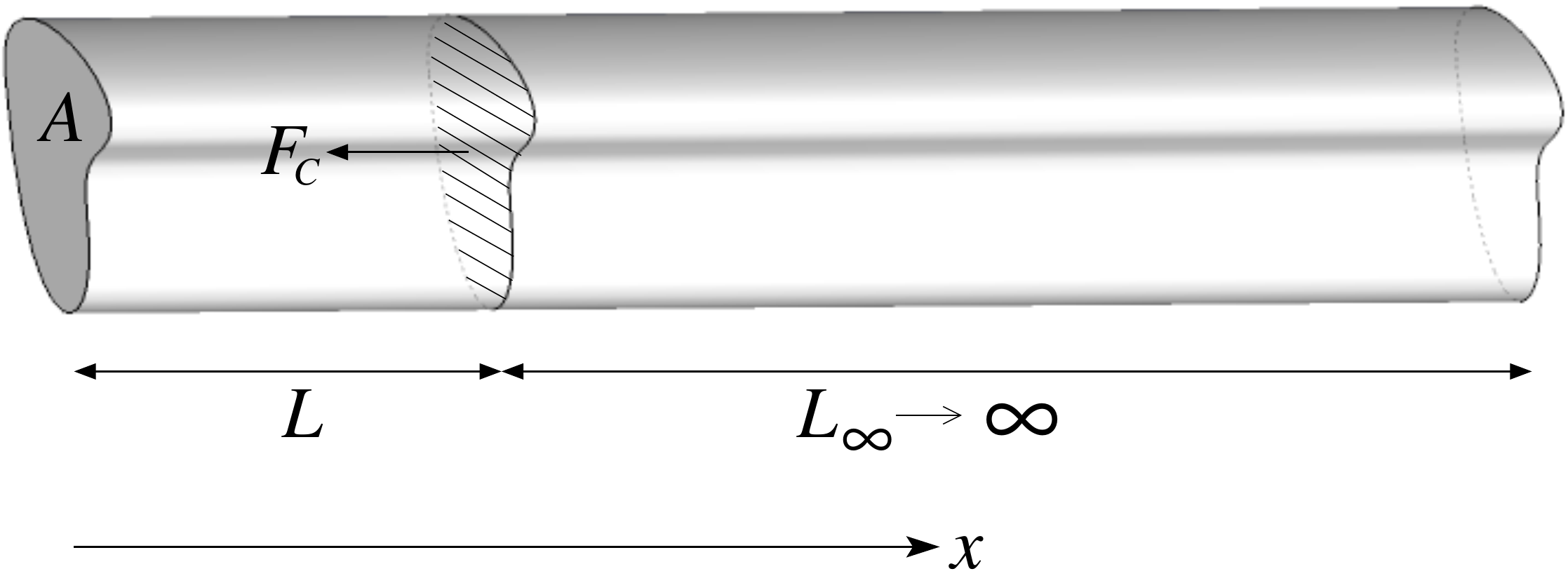}\end{center}
\caption{Geometry of the considered problem. The Casimir force is evaluated for the plate, of area $A$ and general geometry.
The net force is obtained as the force exerted by a plate at distance $L$ and another plate at distance $L_\infty\to\infty$.
The cylinder is oriented along the $x$ axis.} 
\label{fig.plates2}
\end{figure}

\section{Electromagnetic Casimir force for  a piston  of arbitrary cross section} 
\label{sec.em}

\subsection{Electromagnetic Casimir force}
Let us consider the geometry depicted in
Fig.~\ref{fig.plates2}. It consists of a piston of area $A$, but general shape,
made of a perfectly conducting metal surface \cite{Hertzberg}.
Two flat conducting plates of the same cross section of the piston are placed at a~distance~$L$ apart along the $x$ direction.
The plates are perpendicular to the surface of the cylinder.
We calculate the Casimir force, $F_C$, for this plate by evaluating  the expression Eq.~(\ref{eq.T2}) 
for $\langle \Tensor{T}({\bf r}) \rangle$ and later integrating over the surface.
In order to obtain a finite result, we need
a third auxiliary plate located at infinite distance, $L_\infty\to\infty$.

First, we solve the eigenvalue problem for the EM field and apply
Eqs.~\eqref{eq.T} and \eqref{eq.FCint} to obtain the force.
For the EM field the normal component of the stress tensor  reads
$ \mathbb{T}_{xx}=E_{x}^{2} + B_{x}^{2} - \frac{1}{2}\textbf{E}^{2}- \frac{1}{2}\textbf{B}^{2}$,
with BC: $\textbf{E}\times \textbf{n}=\textbf{0}$ and 
$\textbf{B}\cdot\textbf{n}=0$, where $\textbf{n}$ is the surface normal vector.
In this geometry, the EM field can be decomposed in transverse electric (TE)
and transverse magnetic (TM) modes, which are discussed independently~\cite{Jackson}.
For the TM modes, the magnetic field is transverse to the $x$ direction, and
the vector potential ${\bf A}$ can be written as
\beq
{\bf A}=( -C \nabla_\perp^2 D, \partial_y D\, \partial_x C, \partial_zD\,\partial_x C ) e^{-i\omega t}.
\eeq
Here the fields $C(x)$ and $D({\bf r}_\perp)$ satisfy
\begin{eqnarray}\Label{eq.Cg}
\partial_x^2C(x)=&-k_x^2 C(x), &{\text{\ (Neumann BC on\ }} x=0,L{\text{)}} \nonumber  \\
\nabla_{\perp}^{2}D_{n} ({\bf r}_\perp)=& - \lambda_{n}^{2}D_{n}({\bf r}_\perp),  &{\text{\ (Dirichlet BC on  }} {\cal S} {\text{)}},
\end{eqnarray}
where ${\cal S}$ is the surface of the cylinder and ${\bf r}_\perp=(y,z)$.
For the TE set, the electric field is transverse to $x$, so the vector potential is
$ {\bf A}=( 0, -S\,\partial_z N, S\, \partial_yN  ) e^{-i\omega t} $,
where the functions $S(x)$ and $N({\bf r}_\perp)$ satisfy Eqs.~(\ref{eq.Cg})
with the opposite BC: Dirichlet for $S(x)$ and Neumann for $N({\bf r}_\perp)$.
However, in this case, the constant eigenfunction (with eigenvalue $\lambda_n^2=0$)  must be excluded as
it gives ${\bf A}={\bf 0}$ and then ${\bf E}={\bf B}={\bf 0}$.

Substitution of the TE modes into the expression for the stress tensor, and integration over one side of the
plates gives, after a long but straightforward calculation, the Casimir force  as
\begin{equation}\Label{12}
\int_{\tiny{\text{1 side}}}\hspace{-0.4cm}\mean{\mathbb{T}_{xx}^{TE}}
dS_x =
\frac{1}{\beta L}\sum_{m\in\mathbb{Z}}\sum_{n_{x}=1}^{\infty}\sum_{n}
\frac{k_{x}^{2}}{\,\omega_{m}^{2} + k_{x}^{2} + \lambda_{n}^{2}\,},
\end{equation}
where $k_{x}^{2}= (n_x\pi/L)^2$.
For the TM modes, one obtains exactly the same expression, but $\lambda_{n}^2$ are the eigenvalues of the two-dimensional (2D) Laplacian
with Neumann BC. We will denote the complete set of eigenvalues of
the Laplacian with Neumann (excluding the zero eigenvalue)
and Dirichlet BC by the index~$p$.
The expression above is the equivalent of Eq.~\eqref{eq.T} when the spectrum can be split into a~longitudinal and transversal part, that is, $\lambda_n^2= k_x^2+\lambda_p^2$. The sum over the Matsubara frequencies can be performed as in (\ref{eq.T2}). In the limit of vanishing temperature ($\omega_1=2\pi k_B T/\hbar c\to0$) this reduces to 
\begin{equation}\Label{12c}
\lim_{T\to0}
\int_{\tiny{\text{1 side}}}\hspace{-0.4cm}\mean{\mathbb{T}_{xx}^{{\rm total}}}
dS_x
= \hbar c\pi^2\sum_{n_{x}=1}^{\infty}\sum_{p}
\frac{n_{x}^{2}}{L^3\sqrt{n_{x}^{2}\pi^2/L^2 + \lambda_{p}^{2}}}.
\end{equation}

\subsection{Regularization}
The series (\ref{12c}) and the one  at finite temperature are divergent, but the net Casimir force,
which is the difference between the force exerted by the plate at distance $L$, and
the plate at $L_\infty\to\infty$ (see Fig.~1), is finite. To apply the regularization method described in Section \ref{subsec.regularization}, an eigenvalue ordering method is used to obtain an equivalent of total force (on both sides) of each mode. That is, a kernel ordering method is considered. In the case of vanishing temperature this reads
\beq
F_{T=0} = \lim_{Q\to\infty}
\hbar c\pi^2
\sum_{n_{x}=1}^{\infty}\sum_{p} n_x^2
\left[ \frac{K\hspace*{-0.5mm}\left[(\textstyle{\frac{n_{x}^{2}\pi^2}{L^2}} 
+ \lambda_{p}^{2})/Q\right]}{L^3\sqrt{\frac{n_{x}^{2}\pi^2}{L^2}+ \lambda_{p}^{2}}}
-
\frac{K\hspace*{-0.5mm}\left[(\textstyle{\frac{n_{x}^{2}\pi^2}{L^2_\infty}}+ \lambda_{p}^{2})/Q\right]}
{L_\infty^3\sqrt{\frac{n_{x}^{2}\pi^2}{L^2_\infty}+ \lambda_{p}^{2}}}
 \right], \label{kernel.regularization}
\eeq
where $K(x)$ is the regularizing kernel that must satisfy $\lim_{x\to0} K(x)=1$ and $\lim_{x\to\infty} K(x)=0$, 
 and $Q$ is the cutoff eigenvalue. 
 
\begin{figure}[htb]
\begin{center}
\includegraphics[width=0.8\columnwidth,angle=0]{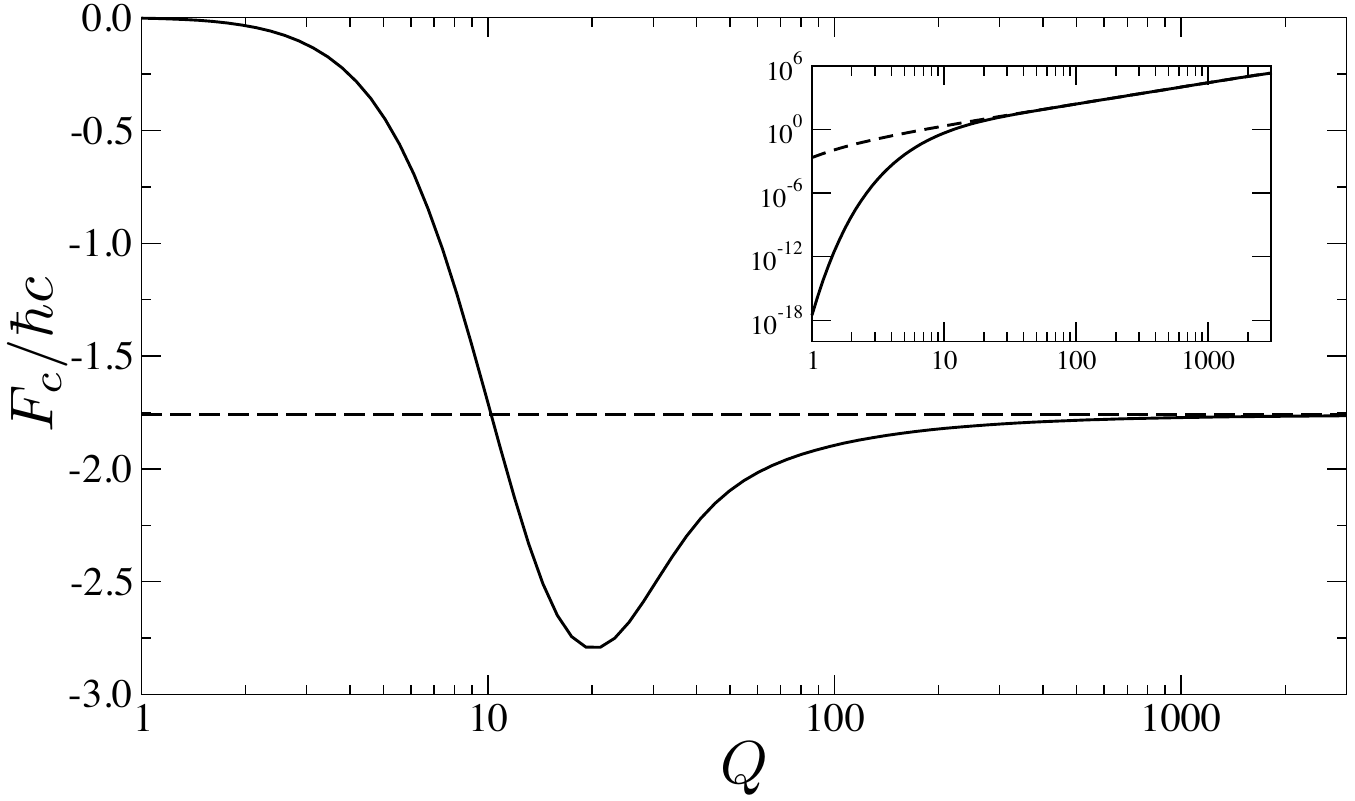}
\end{center}
\caption{Convergence of the Casimir force on a circular plate using the kernel regularization method with $K(x)=e^{-x}$. The circle radius is $R=1$, the interior plate distance is $L=0.5$ and the exterior plate distance is $L_\infty=100.0$. The number of eigenvalues used in the evaluation  is 3000, being $\lambda^2_{3000}\approx77.5$.  Main figure: net force (\ref{kernel.regularization}) as a function of the cutoff value $Q$ (solid line) and the asymptotic value obtained using the analytic regularization (\ref{eq.FCBessel}) (dashed line). Inset: force on the interior  (solid line) and exterior (dashed line) sides of the plate as a function of $Q$.}
\label{fig.kernel}
\end{figure}

To study the convergence of the kernel regularization (\ref{kernel.regularization}) we consider the case of a circular plate. In this case, the eigenvalues are the zeros of the Bessel function $J_\nu(r)$ 
and its derivative. The sum can be done numerically and Fig.~\ref{fig.kernel} shows $F_C$ as a function of $Q$. It is observed that although the force on both sides diverge when $Q\to\infty$ (see inset of Fig.~\ref{fig.kernel}), the net force on the plate converges to a finite value.

For a general piston geometry, the force can be regularized analytically using the Chowla--Selberg summation formula to rewrite the sum over the variable $n_x$ in Eq.~(\ref{12})~\cite{Elizalde}. This formula extracts the divergent, $L$-independent
part of the summation, which cancels when the integral in Eq.~(\ref{12}) is
performed for both sides of the piston. Equivalently, the $L$-independent contribution is the same for
the second plate at distance $L$ or at distance $L_\infty$,  resulting in
\beq\Label{FT}
F_C = - \frac{1}{\beta}\sum_{p}\sum_{m\in \mathbb{Z}} \frac{\sqrt{m^2 \Lambda^2+\lambda_p^2}}{e^{2 L \sqrt{m^2 \Lambda^2+\lambda_p^2}} - 1}.
\eeq
Here, $\Lambda=2\pi /\beta\hbar c $ is the inverse thermal wavelength. This expression gives the finite or regularized Casimir force between two plates
at distance~$L$, valid for any cross section and temperature. The precise geometry of the plates enters into the double
set of eigenvalues of the Laplacian 
$\lambda_p^2$.

In the limit of vanishing temperature the sum over~$m$ in Eq.~(\ref{FT}) 
can be rigorously replaced by an integral, resulting in 
\beq\Label{eq.FCBessel}
 F_{T=0}= - \frac{\hbar c}{2\pi}
 \sum_{p}\sum_{n=1}^\infty \lambda_p^2   \left[ K_0(2n L \lambda_p )+
  K_2(2n L \lambda_p)\right] .
 \eeq
Here $K_{\alpha}(x)$ is the modified Bessel function of order~$\alpha$.
In Ref.~\refcite{Marachevsky} it was obtained a formula for the free energy of the configuration considered here, that, after differentiation with respect to the distance between the plates, leads to 
the force above. 

In a similar fashion, we can calculate the thermal Casimir force when $\hbar \to 0$. Then, in Eq. (\ref{FT}), only the term with $m=0$
is different from zero, resulting in
\begin{equation}\Label{eq.ant}
F_{\hbar=0} = - \frac{1}{\beta}\sum_{p}\frac{\lambda_{p}}{e^{2L\lambda_{p}} - 1}.
\end{equation}

\subsection{Short  and long distances}
For short distances, the summation over $p$ can be calculated
without explicitly knowing the shape of the piston and its eigenvalues. For simple cross sections, 
the eigenvalues, $\lambda_p^2$,  scale with the inverse of the square of the typical size of the piston because
of dimensional arguments. 
So, for
distances~$L$ much smaller than the section of the piston, we can
replace the sum over the eigenvalues, $\lambda_p^2$, by an integral.
In doing that step, we need the 
asymptotic expression for the density of states of the
 Laplacian in two dimensions, that reads (for both Dirichlet or Neumann BC~\cite{DoS}):
 \beq
 \rho_{\eta}(\lambda) = \left[2\lambda A + \eta P\right] \theta(\lambda)/4\pi + \chi\delta(\lambda),
 \eeq
 where $A$ is the piston area, $P$ its diameter, $\eta=1 (-1)$ for Dirichlet (Neumann) and $\chi$ is a measure of the perimeter curvature defined as follows. For a piecewise continuous boundary, 
 \beq
 \chi = \frac{1}{24}\sum_{i}\left[\frac{\pi}{\alpha_{i}} - \frac{\alpha_{i}}{\pi}\right] + \frac{1}{12\pi}\sum_{j}\int_{\gamma_{j}}\kappa(\gamma_{j})d\gamma_{j},
\eeq
where $d\gamma_{j}$ are the line elements of the continuous segments,   $\kappa(\gamma_{j})$ the finite curvature at each point of the boundary, and $\alpha_{i}$ is the angle of each vertex joining contiguous segments. 

When integrating using the density of states, the zero eigenvalue of the Neumann BC  must be excluded. Therefore, the expression (\ref{FT})  in the limit of short distances can be computed as
\begin{align*}
F_{C} =& - \frac{1}{\beta} \sum_{\eta=-1,1}\int_0^\infty  d\lambda \,\rho_{\eta}(\lambda)\sum_{m\in \mathbb{Z}} \frac{\sqrt{m^2 \Lambda^2+\lambda^2}}{e^{2 L \sqrt{m^2 \Lambda^2+\lambda^2}} - 1}  + \lim_{\lambda\to0} \frac{1}{\beta} \sum_{m\in \mathbb{Z}} \frac{\sqrt{m^2 \Lambda^2+\lambda^2}}{e^{2 L \sqrt{m^2 \Lambda^2+\lambda^2}} - 1},
\end{align*}
where the total density of states (Dirichlet plus Neumann) has been considered. After a lengthy but straightforward calculation
\begin{align}
F_{C} =& - k_{B}T\frac{A}{4\pi L^{3}}{\sum_{m=0}^{\infty}}'\left[{\rm Li}_{3}(e^{-2Lm\Lambda}) + 2Lm\Lambda {\rm Li}_{2}(e^{-2Lm\Lambda}) + 2L^{2}m^{2}\Lambda^{2}{\rm Li}_{1}(e^{-2Lm\Lambda})\right] \nonumber\\
& - 2k_{B}T(2\chi - 1)\sum_{m=1}^{\infty}\frac{m\Lambda}{e^{2 L m\Lambda} - 1},
\end{align}
where the apostrophe in the sum indicates that the term $m=0$ should be multiplied by a factor $1/2$ and ${\rm Li}_{s}(z)$ are polylog functions. Note that the force depends on the area and the curvature, but not on the perimeter of the piston.

Taking the limit of low temperatures the quantum Casimir force is
\begin{equation}\Label{eq.NearT0}
F^{\rm near}_{T=0} = - \frac{\hbar c A\pi^{2}}{240 L^{4}} - \frac{\hbar c\left(2\chi - 1\right)\pi}{24L^{2}}.
\end{equation}
The first term is the well-known result of the EM Casimir force for infinite  parallel plates~\cite{Casimir_Placas_Paralelas} and the second term is a contribution of the curvature. 

In the thermal case, of high temperatures, the force  only depends on the area,
\begin{equation}\Label{eq.Nearh0}
F^{\rm near}_{\hbar=0} = -  \frac{k_{B}TA}{4L^{3}}\zeta(3).
\end{equation}

In the opposite  long distance limit, when~$L$ is much larger than the typical size of
the plate only the smallest eigenvalue~$\lambda_{1}^2$ contributes to the sum, with the result
\begin{align}
F^{\rm far}_{T=0} = - \frac{\hbar c}{2\sqrt{\pi L}}g_{1}
\lambda_{1}^{3/2}e^{-2L\lambda_{1}} , \quad
F^{\rm far}_{\hbar=0} = - k_BTg_{1}\lambda_{1}e^{-2L\lambda_{1}} \Label{eq.Far},
\end{align}
being $g_{1}$ the degeneracy of $\lambda_{1}$.

\section{Polygonal pistons} \label{sec.polygons}
A particularly interesting case is that of pistons with a regular polygonal cross section. If $N$ is the number of sides of the polygon and $R_c$ is the exterior (circumscrit) radius of it, the polygon area and curvature parameter are
\begin{align}
A &= \frac{\sin(2\pi/N)}{2\pi/N} \pi R_c^2 , \quad \chi = \frac{N-1}{6(N-2)}.
\end{align}
Using the value of the first eigenvalue, the far and near regimes of the force can be easily computed.

To illustrate the intermediate behavior and the transition from the near to the far regime, we study the case of a equilateral triangle of side $a$. The relevant quantities for the computation in the asymptotic regimes are \cite{triangle}
\begin{align}
A=a^2\sqrt{3}/4;\quad
\chi=1/3;\quad
\lambda_1^2 = 16\pi^2/9a^2\, {\rm (Neumann)};\quad 
g_1 = 2.
\end{align}
At intermediate distances, that is, $L$ comparable to the size of the plates, one must
solve the eigenvalue problem which can be obtained numerically using finite element methods for a general geometry or analytically for special geometries as the present case \cite{triangle}. The dependence with the side of the triangle is 
$\lambda_p^2(a)=\lambda_p^2(a=1)/a^2$, so in the quantum limit the Casimir force as expressed in Eq. (\ref{eq.FCBessel}) is a function of $L/a$ when the force is multiplied by $a^2$, while in the thermal limit  the Casimir force Eq. (\ref{eq.ant}) is a function of $L/a$ when the force is multiplied by $a$.

\begin{figure}[htb]
\begin{center}
\includegraphics[width=0.48\columnwidth,angle=0]{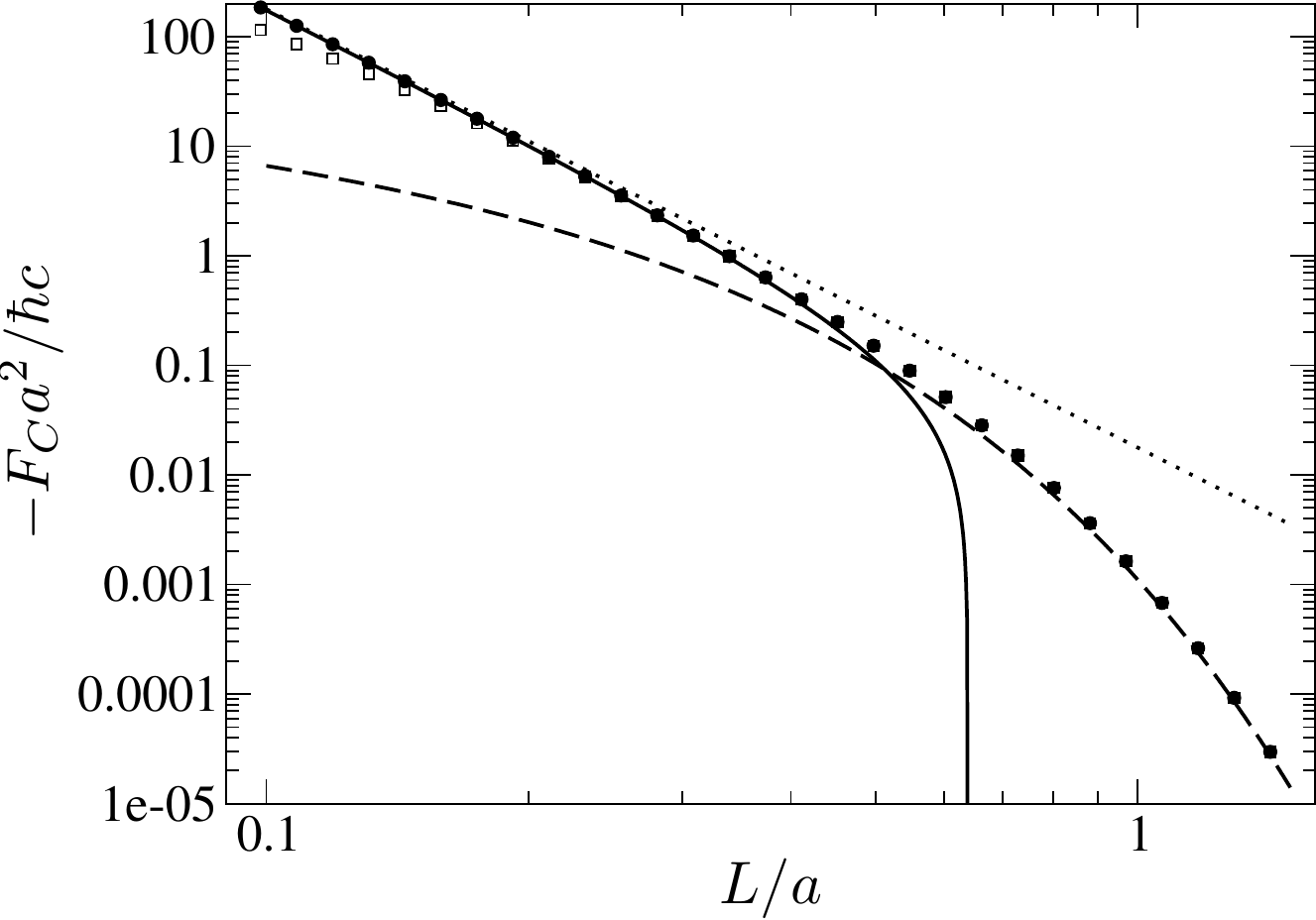} \  \ 
\includegraphics[width=0.49\columnwidth,angle=0]{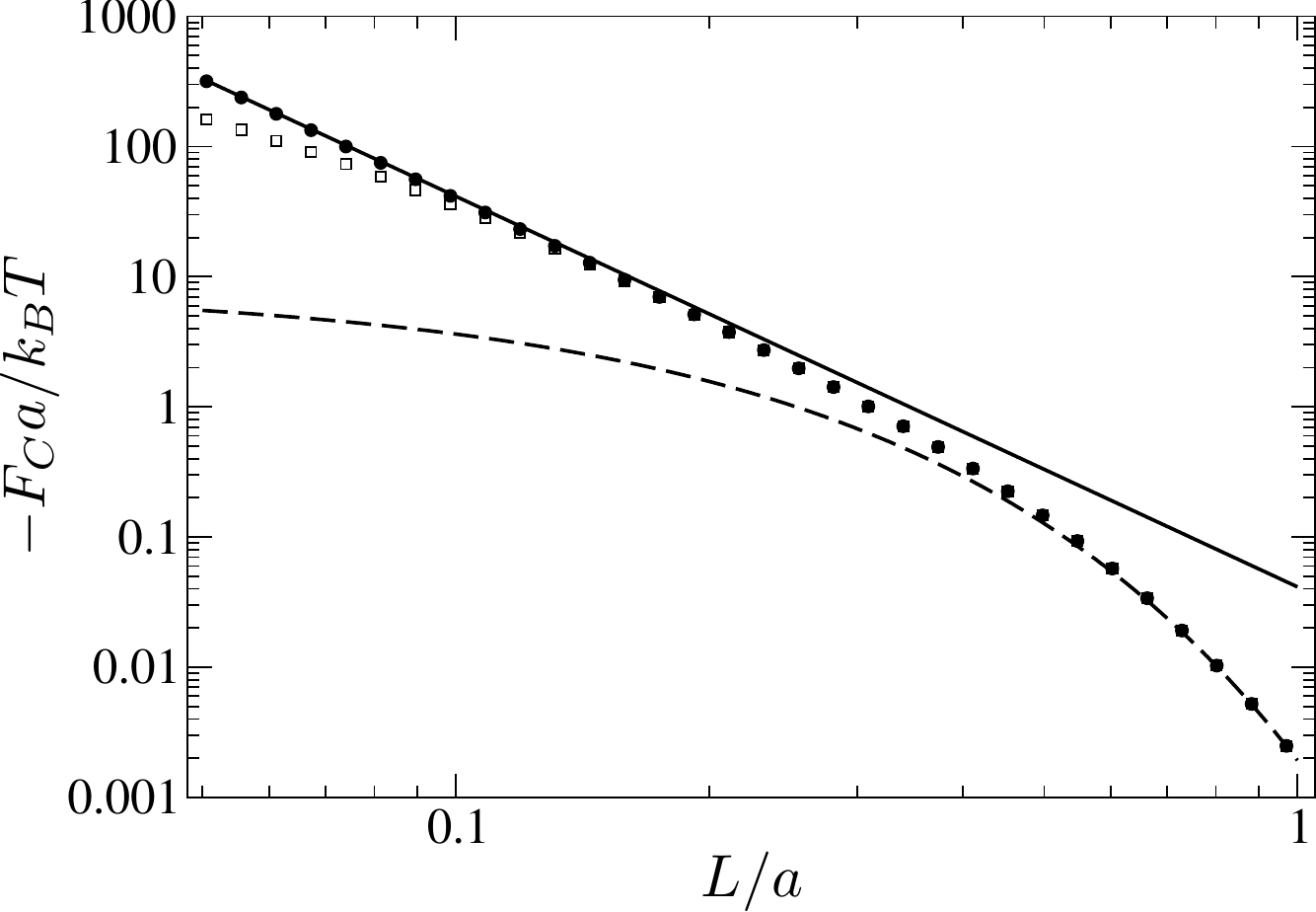}
\end{center}
\caption{{\bf Left:} Quantum limit of the Casimir force for a  piston of triangular cross section of side $a$ as a function of the distance~$L$ between the plates. Solid circles are obtained summing over 4000 eigenvalues of the Laplacian, using Eq.~(\protect{\ref{eq.FCBessel}}). Open squares (overlapping with circles for $L/a>0.2$) are obtained summing only over 20 eigenvalues. Asymptotic expressions for near plates (\ref{eq.NearT0}) without the curvature correction (dotted line), with the curvature correction (solid line) and the asymptotic expression for far plates (\ref{eq.Far}) (dashed line).
{\bf Right:} Thermal limit of the Casimir force for the same configuration.
Solid circles are obtained summing over 4000 eigenvalues of the Laplacian, using Eq.~(\protect{\ref{eq.ant}}). Open squares are obtained summing only over 40 eigenvalues. Asymptotic expressions for near plates (\ref{eq.Nearh0})  (solid line) and the asymptotic expression for far plates (\ref{eq.Far}) (dashed line).}
\label{fig.triangle}
\end{figure}

Figure \ref{fig.triangle} (left panel),  shows  the Casimir force in the quantum limit, Eq. (\ref{eq.FCBessel}), for an equilateral triangle of side $a$. To compare, five results are presented: (1) the full computation using an extensive list of eigenvalues ($N=4000$, that ensures convergence for $L /a\geq0.03$) that we consider as the true value; (2, 3)  the asymptotic expression for near plates (\ref{eq.NearT0}) with and without the curvature correction; (4) the asymptotic expression for far plates (\ref{eq.Far}); (5) a partial sum considering only the first 20 eigenvalues. A similar comparison in the thermal limit is presented in the same figure (right panel).
The analysis of both cases, quantum and thermal, is qualitatively similar. When the plates are separated, only a small number of eigenvalues is needed to obtain the force (20 eigenvalues for $L/a>0.2$). At smaller distances, more eigenvalues would be needed, but there is an excellent match with the asymptotic expansion at short distances. The curvature correction in the quantum limit allows to extend its validity up to $L/a\approx 0.4$.

\section{Conclusions} \label{sec.conclusions}
The new method described in Ref.~\refcite{ParisiWuCasimir} to compute Casimir forces using the stochastic quantization formalism has been used to calculate the force on pistons 
of arbitrary cross section. The method is quite simple and it provides  the force directly, instead of the free energy.
It only requires spectral decomposition of the Laplacian
operator in the given geometry, summation of the eigenvalues and the integration of the eigenfunctions
along the boundary of the object. Such integration of the eigenfunctions over the surface of the body leads to a regularization of
the Casimir force, producing finite results for the force. Quantum and classical limits are recovered as well as intermediate results for finite temperatures.
The electromagnetic Casimir force on  pistons with arbitrary cross section is analyzed in detail. The computation of the 2D eigenvalue problem allows to obtain the Casimir force in the whole range of separation distances. To demonstrate the method, the force on a piston of triangular cross section is computed numerically.  

P.R.-L. and R.B. are supported by the Spanish projects
MOSAICO, FIS2011-22644 
 and MODE\-LICO.
 P.R.-L.'s research is also supported by an FPU MEC grant.
R.S. is supported by Fondecyt grant 1100100
and Proyecto Anillo ACT 127.

\end{document}